\documentstyle[11pt,moriond,epsfig]{article}                                                                         
                                                                         
\def\lapproxeq{\lower .7ex\hbox{$\;\stackrel{\textstyle                                                                         
<}{\sim}\;$}}                                                                         
\def\gapproxeq{\lower .7ex\hbox{$\;\stackrel{\textstyle                                                                         
>}{\sim}\;$}}                                                                         
\def\be{\begin{equation}}                                                                         
\def\ee{\end{equation}}                                                                         
\def\bea{\begin{eqnarray}}                                                                         
\def\eea{\end{eqnarray}}

\begin{document}                                                                         
\vspace*{4cm}                                                                         
                                                                                                                                                 
\title{PENETRATION OF THE EARTH BY ULTRAHIGH ENERGY NEUTRINOS AND THE PARTON
DISTRIBUTIONS INSIDE THE NUCLEON}                                                                         
                                                                         
\author{J.~KWIECINSKI$^{a,b}$,A.D.~MARTIN$^{b}$,A.M.STASTO$^{a,b}$}
\address{$^a$ H.~Niewodniczanski Institute of Nuclear Physics, ul.~Radzikowskiego 152,                                                             
Krakow, Poland \\                                              
$^b$ Department of Physics, University of Durham, Durham, DH1 3LE }                                                                      
                                                                         
                                                                                                                                          \maketitle\abstracts{                                                                       
In this talk we would like to present recent calculations of the cross sections for the ultrahigh energy neutrinos interacting with nucleons. We briefly present the framework of unifed BFKL/DGLAP equations
which resums all the leading $\log(1/x)$ effects as well as the 
subleading terms. The few free parameters which specify the                                              
input parton distributions are determined by fitting to HERA deep inelastic 
data. We then use these parton distributions to calculate the cross section 
for the $\nu N$ interactions at very high energies, up to $10^{12} \; {\rm GeV}$.
 We do also investigate the attenuation of neutrinos
when traversing through the Earth. We use the transport equation which
we solve for different fluxes of high energy neutrinos (Active Galactic
Nucleai, Gamma Ray Bursts, and top-down models). We study the effects 
of the regeneration of the neutrino flux via neutral current interaction.}

The ultrahigh energy neutrinos of energies reaching $10^{12} 
\; {\rm GeV}$ can interact very strongly with nucleons. This implies that they
can be very strongly attenuated by the matter in Earth. Therefore
the exact and detailed knowledge of the value of the neutrino-nucleon
 cross sections plays a vital role for the observations of high energy 
cosmic rays in experiments like Amanda, Nestor, Pierre Auger Cosmic Ray observatory.
These cross sections require the knowledge of parton distributions at small
values of Bjorken $x$.
Normaly these distributions are determined by the HERA and fixed target data
on deep inelastic scattering and they follow standard DGLAP evolution.
However in the region of ultrahigh energies one probes very small values
of the Bjorken $x$ parameter, down to $x \sim 10^{-8}$ or so.
This raises the question whether the $\log(1/x)$ effects could
play an important role. The BFKL equation \cite{BFKL} performs such resummation
up to NLO. However it also poses some problems because the exact
solution for the hard pomeron intercept occurs to be unstable.
 In order to calculate the parton distributions inside the nucleon we shall use here the framework of unifed BFKL/DGLAP evolution
equations \cite{KMS} which treats leading $\log(Q^2)$ and $\log(1/x)$ terms on equal footing. It also resums a major part of the subleading
effects in $1/x$ leading to a stable solution for a pomeron intercept,
yet of a much lower value than the leading one.
 The basic quantity here is the unintegrated gluon distribution function $f(x,k^2)$ which is connected to the standard integrated one via following relation:
\be
xg(x,Q^2) \; = \; \int^{Q^2} {dk^2 \over k^2} \; f(x,k^2)
\label{eq:a1}
\ee
Here $k^2$ is the value of gluon momentum at the end of the ladder.
The observable structure functions are calculated via so called high
energy factorisation theorem:
\be
F(x,Q^2) \; = \; \int {dz \over z} \int {dk^2 \over k^2}
F^{box} (z, k^2,Q^2) f(x/z,k^2)
\label{eq:a2}
\ee
where $F^{box} (z, k^2,Q^2) $ is the partonic cross section for the 
$g\gamma^* \rightarrow q\bar{q}$ process.
Using these two prescriptions we can write the unifed BFKL/DGLAP 
equations in the following symbolic way,
\bea
f(x,Q^2) & = & f^{(0)} \; + \; K^{BFKL}(z,k^2) \otimes f({x \over z},k^2) \; + \; (P_{gg}-1) \otimes 
f({x \over z},k^2) \; + \; P_{gq} \otimes_z 
\Sigma({x \over z},Q^2) \nonumber \\
\Sigma(x,Q^2) & = & \Sigma^{(0)} \; + \; F^{BOX}(x,z,k^2) \otimes f({x \over z},k^2) \; + \;
 P_{qq} \otimes_z \Sigma(z,Q^2)
\label{eq:a3}
\eea
$f(x,k^2)$ and $\Sigma(x,k^2)$ are the gluon and the singlet quark distribution, the two unknown
functions. $f^{(0)}$ and $\Sigma^{(0)}$ are two input distributions which we
parametrise. In the first equation
above we have included BFKL part which is denoted as $K^{BFKL}$ and the
 DGLAP part:
the splitting functions $P_{gg}$ and $P_{gq}$. The $-1$ means here that we have to subtract the leading
term in $1/z$ which is already included in the BFKL part. Second equation includes the high energy factorisation
and also the remaining DGLAP part in $P_{qq}$ function.
$\otimes$ means the convolution in both $k^2$ and $z$ whereas $\otimes_z$ 
denotes the convolution only in $z$ variable. The BFKL part is in principle
in leading order but it also contains very important subleading effects which
are brought by imposition of the consistency constraint. 
This  constraint comes from the fact that the virtuality of the momenta of the exchanged
gluons are dominated by their transverse parts.
 The resulting structure function $F_2$ is fitted to the data  from HERA by changing
available free parameters in $f^{(0)}$ and $\Sigma^{(0)}$. 
 Using these resulting parton distributions we are able to calculate the cross section for the
interaction of neutrino with the nucleon via neutral and charged current interactions.
These results are presented in Fig.1 where we have plotted the cross section up to
energies $10^{12} \; {\rm GeV}$. This result is very close to the other predictions
based on GRV \cite{GKR} and CTEQ4M \cite{GQRS1} partons. We can thus conclude that the difference in extrapolating
the cross sections is less than it might have been expected previously. We see also,
that the cross section grows very fast therefore we expect that
the interactions with matter in Earth can be very important. To study this we use the transport
equation \cite{NIK} of the following form,
\be
{dI(E,\tau)\over d\tau} \; = \; - \sigma_{\rm TOT}(E) I(E,\tau) \: + \: \int {dy\over 1-y}                                    
{d\sigma_{\rm NC}(E^{\prime},y) \over dy} I(E^{\prime},\tau)                                   
\label{eq:a4}
\ee
where $\sigma_{\rm TOT} = \sigma_{\rm CC} + \sigma_{\rm NC}$ and where $y$                                    
is, as usual, the fractional energy loss such that                                   
\be                                   
\label{eq:a5}                                   
E^{\prime}={E\over 1-y}.                                   
\ee                                  
The variable $\tau$ is the number density of nucleons $n$ integrated along a path of                                    
length $z$ through the Earth                                   
\be                                   
\label{eq:a6}                                   
\tau \; = \; \int_0^z \: dz^\prime \: n (z^\prime).                                   
\ee                 
The number density $n(z)$ is defined as $n(z)=N_A \rho (z) $ where $\rho (z)$ is the density of Earth along the neutrino path length $z$ and $N_A$ is the Avogadro number.                  
\begin{figure}[ht!]
\centerline{\epsfig{file=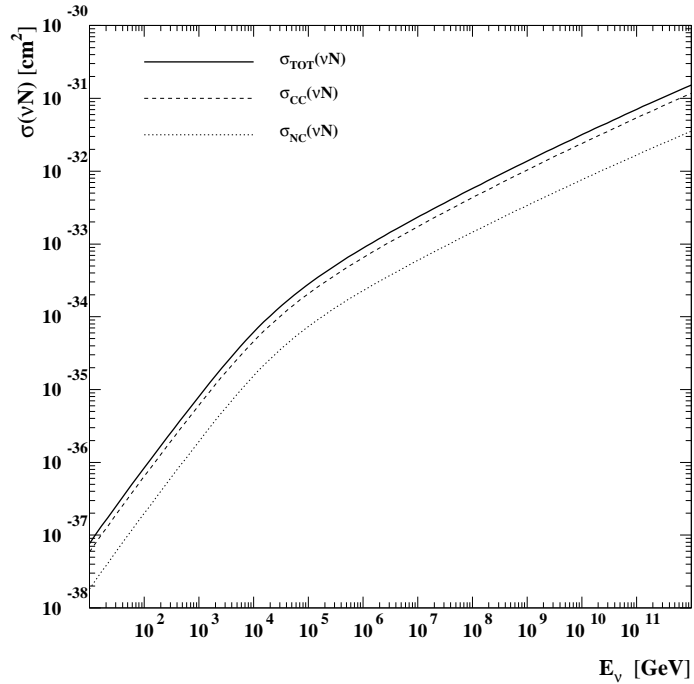,height=10cm,width=10cm}}
\caption{The total $\nu N$ cross section together with its charged 
current and                      
neutral current components as a function of the laboratory neutrino energy.}
\label{fig:fig1}
\end{figure}
We have solved this equation for a variety of initial fluxes $I_0(E)=I(E,\tau)$ coming from Active Galactic Nucleai
Gamma Ray bursts and top-down models. The results for three different models of AGN fluxes
\cite{AGN1,AGN2,AGN3} are compared on Fig. \ref{fig:fig2} 
with the atmospheric background.
They are plotted  as  function of energy for different nadir angles 
(i.e. angle between the normal
to the Earth's surface and the direction of the neutrino beam incident 
into the detector.
We observe that the attenuation of neutrinos is important for neutrinos 
with very high energies $> 10^{8} \; {\rm GeV}$
or so. We have also found that the effect of the neutrino regeneration via 
neutral current interaction
(second term in eq. (\ref{eq:a4}) ) plays an important role for the flat 
spectra of energies and large paths.
This phenomenon is caused by the fact that the neutrinos at certain energies
 are regenerated via neutral
current interactions at lower energies.
This effect can be as high as 50 \% for the fluxes from AGN. \\
\newpage
To sum up we have demonstrated that a framework which incorporates QCD  
expectations at low $x$, including the BFKL effects with the resummation                  
of the non-leading $\ln(1/x)$ terms, gives neutrino cross sections which are                
compatible                 
with those obtained within the NLO DGLAP framework.  This strongly                  
limits potential ambiguities in the possible values of the cross sections                  
extrapolated  from the HERA domain to the region of $x$ and $Q^2$ which can be                  
probed in ultrahigh energy neutrino interactions. Due to large values of these                  
cross sections the attenuation effects reduce the fluxes of ultrahigh                  
energy neutrinos particularly at small nadir angles.  Nevertheless there is a window for  
the observation of AGN by km$^3$ underground detectors of the energetic decay  
muons.  We have found that the AGN flux             
exceeds the atmospheric neutrino background for neutrinos energies $E \gapproxeq             
10^5$~GeV.  Typical results are shown in Fig.~\ref{fig:fig2}.  These illustrate the possibility of  
observing AGN at various nadir angles by \lq\lq neutrino astronomy\rq\rq. \\                
  The details of the calculation as well as results for the other fluxes
  are given in \cite{KMS2}.
\begin{figure}[!ht]
\centerline{\epsfig{file=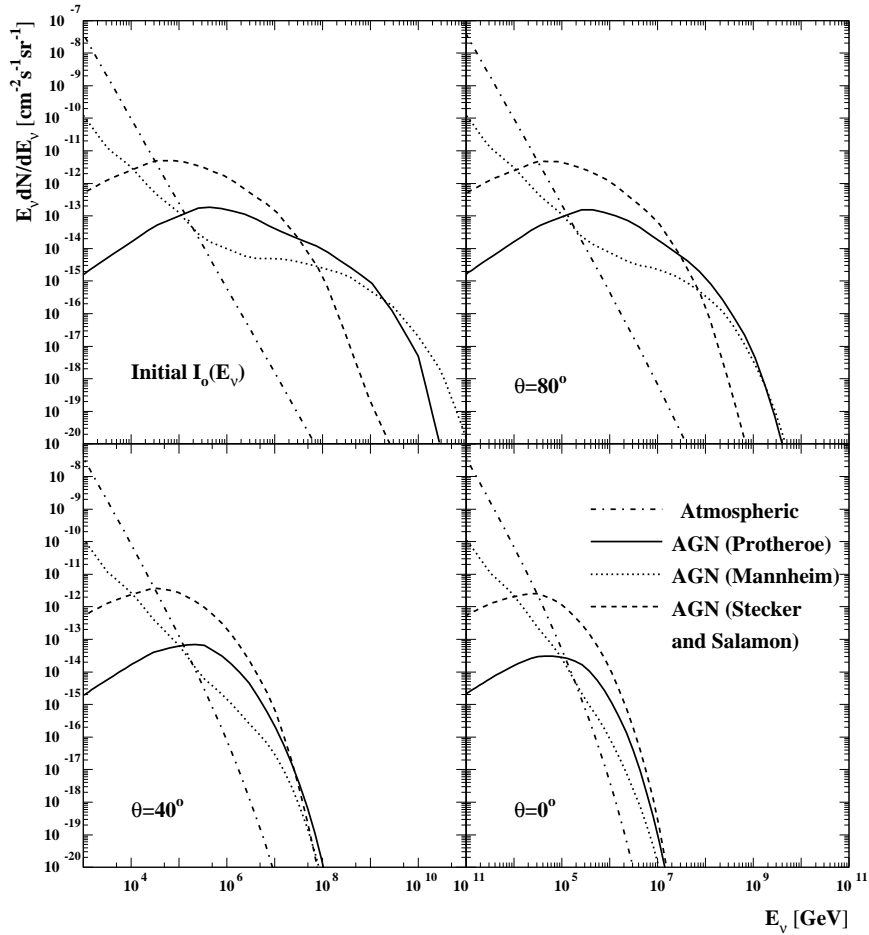,height=13cm,width=13cm}}
\caption{ The initial flux $I_0 (E)$ and the flux at the detector $I (E)$ for three                      
different nadir angles corresponding to three models for AGN neutrinos.                 
 The background atmospheric neutrino flux is also                 
shown.  All the fluxes are given for muon neutrinos. }
\label{fig:fig2}
\end{figure}
\section*{Acknowledgments}
This research has been supported in part by the                 
Polish State Committee for Scientific Research (KBN) grant N0~2~P03B~89~13 
and by the EU Fourth Framework Programme \lq\lq Training and                 
Mobility of Researchers", Network \lq\lq Quantum Chromodynamics and the Deep                 
Structure of Elementary Particles", contract FMRX - CT98 - 0194.  AMS also thanks
Foundation for Polish Science for support.


\begin{thebibliography}{99}
\bibitem{BFKL} E.A.\ Kuraev, L.N.\ Lipatov and V.S.\ Fadin, Sh.\ Eksp.\ Teor.\                                            
Fiz.~{\bf 72} (1977) 373, (Sov.\ Phys.\ JETP~{\bf  45} (1977) 199); Ya.\ Ya.\                                             
Balitzkij and L.N.\ Lipatov, Yad.\ Fiz.~{\bf 28} (1978) 1597 (Sov.\ J.\ Nucl.\                                           
Phys.~{\bf 28} (1978) 822), J.B.\ Bronzan and R.L.\ Sugar, Phys.\ Rev.~{\bf D17}                                           
(1978) 585; T.\ Jaroszewicz, Acta.\ Phys.\ Polon.~{\bf B11} (1980) 965.
\bibitem{KMS} J.~Kwiecinski, A.D.~Martin and A.M.~Stasto, Phys.~Rev.~{\bf                                           
D56} (1997) 3991; \\                               
J.~Kwiecinski, A.D.~Martin and A.M.~Stasto, Acta Phys.~Polon. {\bf B28, No.12}                 
(1997) 2577.                            
%
\bibitem{GKR} M. Gl\"uck, S. Kretzer and E. Reya, DO-TH-98-20, {\tt                             
astro-ph/9809273}.        
\bibitem{GQRS1} R. Gandhi, C. Quigg, M. Reno and I. Sarcevic, Phys. Rev. {\bf                             
D58} (1998) 093009.                            
                                     
\bibitem{NIK} A.~Nicolaidis and A.~Taramopoulos, Phys.Lett.~{\bf B386} (1996)                               
211.                               
\bibitem{AGN1} R. J. Protheroe, Accretion Phenomena and Related Outflows,                            
 IAU Colloq.163, ed. D. Wickramashinghe et al., 1996, {\tt                            
astro-ph/9607165}; Talk at Neutrino 98, Takayama 4-9 June 1998, {\tt                             
astro-ph/9809144}; ADP-AT-96-15, Towards the Millennium                             
in Astrophysics:  Problems and Prospects, Erice 1996, eds. M.M. Shapiro and J.P.                             
Wefel (World Scientific, Singapore), {\tt astro-ph/9612213}.                            
%
\bibitem{AGN2} K. Mannheim, Astroparticle Physics {\bf 3} (1995) 295.                    
%
\bibitem{AGN3} F.W. Stecker and M.H. Salamon, \lq\lq TeV Gamma-ray                 
Astrophysics''                            
(1994) 341, {\tt  astro-ph/9501064}    
\bibitem{KMS2}  J.~Kwiecinski, A.D.~Martin and A.M.~Stasto, Phys.~Rev.~{\bf                                           
D59} (1999) 093002; \\                          
\end{thebibliography}
\end{document}